\documentclass[12pt]{article}
\usepackage{graphicx}
\usepackage[dvipsname]{xcolor}
\usepackage{amsmath}
\usepackage{amssymb}
\usepackage{currfile}
\usepackage{ulem}
\definecolor{olivegreen}{rgb}{0.3,0.5,0.1}
\definecolor{orange}{rgb}{0.9,0.5,0.0}
\newcommand{\red}[1]{\textcolor{red}{#1}}

\newcommand{\reftitle}[1]{{\it #1}}

\newcommand{\calO}{{\cal O}}

\newcommand{\GeV}{\text{ GeV}}
\newcommand{\TeV}{\text{ TeV}}
\begin{document}
\thispagestyle{empty}
{\small \noindent October 5, 2017    \hfill KEK-TH-1978, OU-HET 934}

\rightline{\small (revised.)}

\vspace{3.0cm}
\baselineskip=35pt plus 1pt minus 1pt
\begin{center}
{\LARGE \bf Distinct signals of the gauge-Higgs unification in $e^+e^-$ collider experiments}
\end{center}
\vspace{1.0cm}
\baselineskip=22pt plus 1pt minus 1pt
\begin{center}
{Shuichiro Funatsu$^1$, Hisaki Hatanaka$^2$, Yutaka Hosotani$^2$ 
and Yuta Orikasa$^3$}
\vskip 10pt

\it\small
$^1$KEK Theory Center, KEK, Tsukuba, Ibaraki 305-0801, Japan\\
\it\small
$^2$Department of Physics, Osaka University, Toyonaka, Osaka 560-0043, Japan\\
\it\small
$^3$Czech Technical University, Prague 12800, Czech Republic\\
\end{center}

\vspace{2.0cm}
\begin{abstract}
Effects of Kaluza-Klein excited neutral vector bosons ($Z'$ bosons)
in the gauge-Higgs unification 
on $e^+e^- \to \bar{q}q, \ell^+\ell^-$ cross sections are studied, 
particularly in future $e^+ e^-$ collider experiments with polarized beams. 
Significant deviations in the energy and polarization dependence in $\sigma(\mu^+\mu^-)$, 
the lepton forward-backward asymmetry, $R_b(\mu) \equiv \sigma(\bar{b}b)/\sigma(\mu^+\mu^-)$ 
and the left-right asymmetry from the standard model are predicted. 
\end{abstract}


\newpage

With the establishment of the standard model (SM) by the discovery of the Higgs boson,
searching for physics beyond the SM and   understanding the electroweak phase transition 
have become a few of the main topics in  particle physics.
Not only  large hadron colliders, but also  $e^+ e^-$ colliders play an important role for this purpose.
In this letter we study distinct signals of the gauge-Higgs unification (GHU) 
\cite{Hosotani:1983xw}-\cite{Funatsu:2016uvi}
in the future $e^+ e^-$ collider experiments.

In GHU the Higgs boson is a part of the extra-dimensional 
component of the gauge potentials, appearing as a fluctuation mode of an 
Aharonov-Bohm (AB) phase $\theta_H$ in the fifth dimension.
As a consequence the Higgs couplings $HWW$, $HZZ$ and 
Yukawa couplings deviate from those in the SM in a universal manner.\cite{HiggsCoupling}
They are suppressed by a common factor $\cos\theta_H$;
\begin{eqnarray}
\frac{g_{HWW}^{GHU}}{g_{HWW}^{\rm SM}},\,
\frac{g_{HZZ}^{\rm GHU}}{g_{HZZ}^{\rm SM}},\,
\frac{y_{\bar{f}f}^{\rm GHU}}{y_{\bar{f}f}^{\rm SM}} \simeq
\cos\theta_H ~.
\label{Hcoupling1}
\end{eqnarray}
For $\theta_H = \calO(0.1)$,  probable values in the model,    
the deviation of the couplings amounts to $1-\cos\theta_H = \calO(0.005)$, and is small.
At the ILC at $\sqrt{s} = 250\GeV$, the $ZZH$ coupling can be measured 
in the $0.6 \%$ accuracy with $2\text{ ab}^{-1}$ data \cite{Barklow:2017suo}.
Another prominent feature of the model is that the first Kaluza-Klein (KK) excited 
states of the neutral gauge bosons,   $Z'$,  have large couplings to right-handed 
components of quarks and leptons,  viable signals of which 
can be seen in hadron collider experiments \cite{Funatsu:2014fda,Funatsu:2016uvi}.

The main purpose of this letter is to check the effect of such $Z'$ bosons using lepton 
collider  experiments in the past and future.
We first examine the GHU model with precision measurements in LEP1 experiment 
at $\sqrt{s} = M_Z$, and LEP2 experiments for $130 \GeV \le \sqrt{s} \le 207 \GeV$.
Then we  predict several  signals of $Z'$ bosons in GHU in  $e^+ e^-$ collider experiments 
designed  for future  with collision energy $\sqrt{s} \ge 250 \text{ GeV}$ with 
polarized electron and positron beams.


The GHU model we consider is  the $SO(5)\times U(1)_X$ gauge theory 
in the Randall-Sundrum warped space  with metric 
$ds^2 = e^{-2k|y|} \eta_{\mu\nu} dx^\mu dx^\nu +dy^2$ ($0 \le |y| \le +L$) where $k$ is the $AdS_5$ curvature. 
The warp factor $z_L \equiv e^{kL}$ is large ($\gg 1$).
$SO(5)$ symmetry is broken to $SO(4) \simeq SU(2)_L \times SU(2)_R$ by the orbifold boundary 
conditions at $y=0$ and $L$.  The $SO(5)/SO(4)$ part of the gauge fields,  
 $A_y^{\hat{a}}$ ($a=1 \sim 4$),  plays the role of the  Higgs field in the SM. 
 $SU(2)_R \times U(1)_X$ symmetry is spontaneously broken 
 to $U(1)_Y$ by  a brane-localized scalar field at $y=0$.
Finally the $SU(2)_L \times U(1)_Y$  symmetry is dynamically broken to 
$U(1)_{\rm em}$ by the Hosotani mechanism.

5D fields are expanded in KK series. In particular,
there are  four KK towers of the neutral vector bosons, $\gamma^{(m)}$, $Z^{(m)}$, 
$Z_R^{(n)}$ and $A^{\hat{4}(n)}$ ($m=0,1,2,\cdots$, $n=1,2,3\cdots$) where 
$\gamma^{(0)}$ and $Z^{(0)}$ correspond to the photon and  $Z$ boson, respectively.
These fields except for $A^{\hat{4}}$ couple to the SM fields and can be observed 
as neutral $Z'$ vector bosons.

In addition to the quark-lepton multiplets in the vector representation of $SO(5)$, 
$N_F$ dark fermions in the spinor representation are introduced.  As a consequence
the  electroweak symmetry breaking is achieved at the one loop level.
The Higgs boson, which is massless at the tree level, acquires a finite mass $m_H$,
independent of the cutoff scale.  The gauge hierarchy problem is thus 
solved.\cite{Hatanaka:1998yp}

There remain two  free parameters, $N_F$ and $z_L$.
Given $N_F$ and $z_L$, the effective potential $V_{\rm eff} (\theta_H)$ is fixed.
From the location of the minimum of $V_{\rm eff} (\theta_H)$, the value $\theta_H$ is determined.
There is the property called the universality such that
many of the physical quantities are determined by $\theta_H$, but do not depend on 
$N_F$ and $z_L$ independently.
In the following we take $N_F=4$ and parameterize the model by $\theta_H$.

We note that some of the composite Higgs models (CHM) have similar features to those 
in the GHU.  In particular, CHM based on $SO(5)$ gauge group has almost the same 
gauge structure as the $SO(5) \times U(1)_X$ GHU \cite{ACP2005, Contino2007}. 
The  $SO(5)$ gauge invariance
is reduced to $SO(4)$ by orbifold boundary conditions in both cases.
However, there are many differences between the two.  
The 4D Higgs boson in GHU is a fluctuation mode of the AB phase $\theta_H$ 
in the fifth dimension, but is not  a pseudo-Nambu-Goldstone boson
supposed in CHM.  Secondly, 
in most of the CHM, $SO(4)$-breaking   boundary conditions are imposed 
on fermion fields by hand to obtain the quark-lepton spectrum.  
In the GHU theory  based on the action principle the $SO(5) \times U(1)_X$ gauge
invariance in the bulk and the $SO(4) \times U(1)_X$ gauge invariance on the UV
and IR branes are strictly preserved.  
GHU is more restrictive than CHM, and is powerful to make predictions.

In GHU the relevant parameter for physics of SM particles is $\theta_H$.  
With $\theta_H$ given, the KK spectra of various fields, the couplings of 
quarks and leptons to KK gauge bosons, and the Higgs couplings are all determined.
The Higgs boson mass $m_H \sim 125\,$GeV and $m_{KK} = 7 \sim 10\,$TeV
are naturally realized for $\theta_H \sim 0.1$ without fine-tuning of the parameters.
It has been shown that corrections to $H \rightarrow \gamma \gamma, Z \gamma$ due to
an infinite number of KK states of $W$, $t$ et al.\ running in the loop 
are finite and tiny for $\theta_H \sim 0.1$.  It has been recognized that the 
$SO(5) \times U(1)_X$ GHU in the RS space gives nearly the same phenomenology
at low energies as the SM for $\theta_H \lesssim 0.1$.

The phase $\theta_H$  in GHU corresponds to the vacuum misalignment angle in CHM.
A bound $\theta_H < 0.3$ has been derived in CHM from the $S$ parameter 
constraint.\cite{Contino2007, Fan:2014vta}
In GHU much stronger constraint $\theta_H \lesssim 0.1$ is obtained from the current 
non-observation of $Z'$ signals at LHC. 
It should be stressed in this connection that in GHU in the RS space right-handed 
quarks and leptons and KK gauge bosons are localized near the IR brane whereas 
left-handed quarks and leptons are localized near the UV brane so that 
right-handed quarks and leptons have larger couplings to KK gauge bosons 
than left-handed quarks and leptons.


%
Masses and widths of $Z'$ bosons are tabulated in Table \ref{tbl:masses}.
Fermion couplings to $Z'$ for $\theta_H=0.115$, $0.0917$ and  $0.0737$
are given in Tables \ref{tbl:coupling_light}, \ref{tbl:coupling_mid} and 
\ref{tbl:coupling_heavy}, respectively.
In the evaluation  $\sin^2\theta_W = 0.23126$ and $M_Z = 91.1876\GeV$ are adopted.
The $Z$ couplings  of quarks and leptons except for top quark are almost the same as in the SM within the accuracy of one part in $10^4$.  
The deviation of the $Zt \bar t$ couplings are
less than 1\%, whereas the deviation of the $Zb \bar b$ couplings  are very tiny in GHU.

\begin{table}[tp]
\caption{Masses and widths of $Z'$ bosons, $Z^{(1)}$, $\gamma^{(1)}$, and
$Z_R^{(1)}$ ($N_F=4$)}
\label{tbl:masses}
\vskip 3pt
\centering
\begin{tabular}{cc|cccccccc}
$\theta_H$ & $\dfrac{z_L}{10^4}$ 
& $m_{KK}$ & $m_{Z^{(1)}}$ &$\Gamma_{Z^{(1)}}$ & 
 $m_{\gamma^{(1)}}$ &$\Gamma_{\gamma^{(1)}}$ & $m_{Z_R^{(1)}}$ & $\Gamma_{Z_R^{(1)}}$
\\ 
{}[rad.] & 
& [TeV] & [TeV] & [GeV] & [TeV] & [GeV] & [TeV] & [GeV] \\
\hline
$0.115$ & $10$ 
& $7.41$ & $6.00$ & $406$& $6.01$ & $909$ & $5.67$ & $729$
\\
$0.0917$ & $3$ 
& $8.81$ & $7.19$ & $467$ & $7.20$ & $992$ & $6.74$ & $853$
\\
$0.0737$ & $1$ 
& $10.3$ & $8.52$ & $564$ & $8.52$ & $1068$ & $7.92$ & $1058$ \\
\end{tabular}
\end{table}

We  evaluate  $e^+e^- \to \bar{f}f$  cross sections $\sigma (\bar f f)$ where $f$ is a lepton or quark. 
In addition to leptonic and hadronic cross sections, 
forward-backward asymmetry defined by
\begin{eqnarray}
A_{\rm FB} &=& \frac{\int_0^1 \frac{d\sigma}{d\cos\theta}d\cos\theta - \int_{-1}^{0} \frac{d\sigma}{d\cos\theta}d\cos\theta}{\int_{-1}^{1} \frac{d\sigma}{d\cos\theta}d\cos\theta},
\end{eqnarray}
the ratio of hadronic and leptonic cross sections 
$R_\mu \equiv \sigma(\bar{q}q)/\sigma(\mu^+\mu^-)$, and
the asymmetry of $\sigma (\bar f f)$ with right- and left-handed polarized electron beams
are investigated.\footnote{In the numerical evaluation in this letter we have 
used the values of the various couplings obtained for $m_H=126$GeV.  
With $m_H=125$GeV  the value of $M_{Z'}$, for instance, decreases by 1.3 \%.}

Cross sections are evaluated to the leading order, which may receive quantum corrections. 
Such corrections are parametrised as
$\sigma \to \delta_{\rm QCD} \cdot \delta_{\rm QED} \cdot \sigma + r_{\rm nf}$ where 
$\delta_{\rm QCD} = 1 + \calO(\alpha_s/\pi)$ and 
$\delta_{\rm QED} = 1 + \calO(\alpha_{EM}/\pi)$ are factorizable QCD and QED corrections,
whereas $r_{\rm nf}$ denotes non-factorizable  corrections.
In this paper we assume
that  $\delta_{\rm QCD,QED}^{\rm GHU} \simeq \delta_{\rm QCD,QED}^{\rm SM}$ 
and $r_{\rm nf}$ for both GHU and SM are small.
We have taken only the first KK states into account. 
The second KK states are approximately twice as heavy as the first KK states.
The magnitudes of couplings of the second KK states are at most a half of 
the couplings of the first KK states.  Thus the contributions of the second KK states are 
expected to be small.

\begin{table}[tbp]
\caption{Couplings of neutral vector bosons ($Z'$ bosons) 
to fermions in unit of $g_w = e/\sin\theta_W$  for $\theta_H = 0.115$.  
Corresponding $Z$-boson coupling in the SM
are $(g_{Z\nu}^L,g_{Z\nu}^R) = (0.57027,0)$,
$(g_{Ze}^L,\,g_{Ze}^R) = (-0.30651,0.26376)$,
$(g_{Zu}^L,\,g_{Zu}^R) = (0.39443,-0.17584)$ and
$(g_{Zd}^L,\,g_{Zd}^R) = (-0.48235,0.08792)$.}
\label{tbl:coupling_light}
\begin{center}
\begin{tabular}{c|cc|cc|cc|cc}
$f$ 
& $g^L_{Zf}$ &  $g^R_{Zf}$ 
& $g^L_{Z^{(1)}f}$ & $g^R_{Z^{(1)}f}$ & $g^L_{Z_R^{(1)}}$ & $g^R_{Z_R^{(1)}f}$ & $g^L_{\gamma^{(1)}f}$ & $g^R_{\gamma^{(1)}f}$ 
 \\
\hline
 $\nu_e$ 
 & $0.57041$ & $0$ & $-0.1968$ & $0$ & $0$ & $0$ & $0$ & $0$ \\
 $\nu_\mu$ 
 & $0.57041$ & $0$ & $-0.1968$ & $0$ & $0$ & $0$ & $0$ & $0$ \\
 $\nu_\tau$ 
 & $0.57041$ & $0$ & $-0.1967$ & $0$ & $0$ & $0$ & $0$ & $0$ \\
\hline
 $e$ 
 & $-0.30659$ & $0.26392$ & $0.1058$ & $1.0924$ & $0$ & $-1.501$ & $0.1667$ & $-1.983$ \\
 $\mu$  
 & $-0.30659$ & $0.26391$ & $0.1058$ & $1.0261$ & $0$ & $-1.420$ & $0.1667$ & $-1.863$ \\
 $\tau$  
 & $-0.30658$ & $0.26391$ & $0.1057$ & $0.9732$ & $0$ & $-1.354$ & $0.1666$ & $-1.767$ \\
 \hline 
 $u$ 
 & $0.39453$ & $-0.17594$ & $-0.1361$ & $-0.7152$ & $0$ & $0.9846$ & $-0.1111$ & $1.2983$ \\
 $c$ 
 & $0.39453$ & $-0.17594$ & $-0.1361$ & $-0.6631$ & $0$ & $0.9205$ & $-0.1111$ & $1.2036$ \\
 $t$ 
 & $0.39339$ & $-0.17712$ & $0.5068$ & $-0.4764$ & $1.0314$ & $0.6899$ & $0.4158$ & $0.8666$ \\
\hline
 $d$ 
 & $-0.48247$ & $0.087972$ & $0.1665$ & $0.3576$ & $0$ & $-0.4923$ & $0.05557$ & $-0.6491$ \\
 $s$ 
 & $-0.48247$ & $0.087970$ & $0.1664$ & $0.3315$ & $0$ & $-0.4602$  & $0.05556$ & $-0.6018$ \\
 $b$ 
 & $-0.48254$ & $0.087964$ & $-0.6303$ & $0.2387$ & $1.0292$ & $-0.3446$ & $-0.2082$ & $-0.4331$ \\
\end{tabular}
\end{center}
\end{table}

\begin{table}[ht]
\caption{$Z'$ couplings of fermions  for $\theta_H = 0.0917$.
Unit is the same as in Table~\ref{tbl:coupling_light}.}
\label{tbl:coupling_mid}
\begin{center}
\begin{tabular}{c|cc|cc|cc|cc}
$f$ & $g^L_{Zf}$ &  $g^R_{Zf}$ & $g^L_{Z^{(1)}f}$ & $g^R_{Z^{(1)}f}$ & $g^L_{Z_R^{(1)}}$ & $g^R_{Z_R^{(1)}f}$ & $g^L_{\gamma^{(1)}f}$ & $g^R_{\gamma^{(1)}f}$ 
 \\
\hline
$\nu_e$      & $0.57037$ & 0 & $-0.2092$ & 0 & 0 & 0 & 0 & 0 \\
$\nu_{\mu}$  & $0.57037$ & 0 & $-0.2092$ & 0 & 0 & 0 & 0 & 0 \\
$\nu_{\tau}$ & $0.57037$ & 0 & $-0.2092$ & 0 & 0 & 0 & 0 & 0 \\
\hline
$e$    
& $-0.30656$ & $0.26387$ & $0.1124$ & $1.0443$ & 0 & $-1.438$ & $0.1769$ & $-1.899$ \\
$\mu$  
& $-0.30656$ & $0.26387$ & $0.1124$ & $0.9804$ & 0 & $-1.361$ & $0.1769$ & $-1.783$ \\
$\tau$ 
& $-0.30656$ & $0.26387$ & $0.1124$ & $0.9278$ & 0 & $-1.296$ & $0.1768$ & $-1.687$ \\
\hline
$u$ 
& $0.39450$ & $-0.17591$ & $-0.1447$ & $-0.6838$ & $0$ & $0.9438$ & $-0.1179$ & $1.2433$ \\
$c$ 
& $0.39450$ & $-0.17591$ & $-0.1477$ & $-0.6328$ & $0$ & $0.8818$ & $-0.1179$ & $1.1505$ \\
$t$ 
& $0.39367$ & $-0.17678$ & $0.5635$ & $-0.4245$ & $1.1239$ & $0.6258$& $0.4606$ & $0.7734$ \\
\hline
$d$ 
& $-0.48243$ & $0.087957$ & $0.1769$ & $0.3419$ & $0$ & $-0.4719$ & $0.05897$ & $-0.6216$ \\
$s$ 
& $-0.48243$ & $0.087955$ & $0.1769$ & $0.3164$ & $0$ & $-0.4409$ & $0.05896$ & $-0.5753$ \\
$b$ 
& $-0.48249$ & $0.087951$ & $-0.6959$ & $0.2127$ & $1.1220$ & $-0.3127$ & $-0.2304$ & $-0.3866$ \\
\end{tabular}
\end{center}
\end{table}

\begin{table}[ht]
\caption{$Z'$ couplings of fermions  for $\theta_H = 0.0737$.
Unit is the same as in Table~\ref{tbl:coupling_light}.}
\label{tbl:coupling_heavy}
\begin{center}
\begin{tabular}{c|cc|cc|cc|cc}
$f$ & $g^L_{Zf}$ &  $g^R_{Zf}$ & $g^L_{Z^{(1)}f}$ & $g^R_{Z^{(1)}f}$ & $g^L_{Z_R^{(1)}}$ & $g^R_{Z_R^{(1)}f}$ & $g^L_{\gamma^{(1)}f}$ & $g^R_{\gamma^{(1)}f}$ 
 \\
\hline
$\nu_e$      & $0.57034$ & 0 & $-0.2225$ & 0 & 0 & 0 & 0 & 0 \\
$\nu_{\mu}$  & $0.57034$ & 0 & $-0.2225$ & 0 & 0 & 0 & 0 & 0 \\
$\nu_{\tau}$ & $0.57034$ & 0 & $-0.2225$ & 0 & 0 & 0 & 0 & 0 \\
\hline
$e$    
& $-0.30655$ & $0.26384$ & $0.1196$ & $0.9981$ & 0 & $-1.376$ & $0.1880$ & $-1.817$ \\
$\mu$  
& $-0.30655$ & $0.26384$ & $0.1196$ & $0.9369$ & 0 & $-1.303$ & $0.1880$ & $-1.705$ \\
$\tau$ 
& $-0.30655$ & $0.26384$ & $0.1195$ & $0.8847$ & 0 & $-1.240$ & $0.1879$ & $-1.610$ \\
\hline
$u$ 
& $0.39448$ & $-0.17589$ & $-0.1539$ & $-0.6536$ & $0$ & $0.9034$ & $-0.1253$ & $1.1896$ \\
$c$ 
& $0.39448$ & $-0.17589$ & $-0.1539$ & $-0.6041$ & $0$ & $0.8439$ & $-0.1253$ & $1.0994$ \\
$t$ 
& $0.39379$ & $-0.17661$ & $0.6888$ & $-0.3431$ & $1.3208$ & $0.5253$ & $0.5616$ & $0.6258$ \\
\hline
$d$ 
& $-0.48241$ & $0.087947$ & $0.1882$ & $0.3268$ & $0$ & $-0.4517$ & $0.06267$ & $-0.5948$ \\
$s$ 
& $-0.48241$ & $0.087946$ & $0.1882$ & $0.3021$ & $0$ & $-0.4320$ & $0.06266$ & $-0.5497$ \\
$b$ 
& $-0.48246$ & $0.087941$ & $-0.8470$ & $0.1720$ & $1.3189$ & $-0.2625$ & $-0.2808$ & $-0.3129$ \\
\end{tabular}
\end{center}
\end{table}


In the LEP1 experiment \cite{ALEPH:2005ab}  at the $Z$-pole ($\sqrt{s} = M_Z$)  
the measured and fitted values of cross sections, forward-backward asymmetries of charged leptons 
$A_{\rm FB}^\ell$, $R_\ell^{0} \equiv \Gamma_{\rm hadrons} / \Gamma_{\rm \ell}$
($\ell = e$, $\mu$) and $R_b \equiv \Gamma_{b}/\Gamma_{\rm hadrons}$ are given by
\begin{eqnarray}
\sigma^{\rm meas}(\bar{q}q) / \sigma^{\rm fit}(\bar{q}q) &=& 
1.00149\pm 0.00089,
\\
A_{\rm FB}^{\ell}{}^{\rm meas} / A_{\rm FB}^{\ell}{}^{\rm fit} &=& 
1.042 \pm 0.058,
\\
R_\ell^{0,\,\rm meas}/ R_\ell^{0,\,\rm fit} &=& 
1.0012\pm 0.0012\red{,}
\\
A_{\rm FB}^b{}^{\rm meas}/A_{\rm FB}^b{}^{\rm fit} &=& 0.956\pm 0.015,
\\
R_b^{\rm meas}/R_b^{\rm fit} &=& 1.002 \pm 0.031,
\end{eqnarray}
where $\sigma^{\rm fit}(\bar{q}q) = 41.478\text{ nb}$,
$A_{\rm FB}^\ell{}^{\rm fit} = 0.01645$,
$R_\ell^0{}^{\rm ,fit} = 20.742$,
$A_{\rm FB}^b{}^{\rm fit}  =0.1038$ and
$R_b^{\rm fit} = 0.21579$.
In GHU, we obtain
\begin{eqnarray}
\sigma^{\rm GHU}(\bar{q}q) / \sigma^{\rm SM}(\bar{q}q) &=& 1.00143,\, 1.00098,\, 1.00073,
\\
A_{\rm FB}^{\ell}{}^{\rm GHU} / A_{\rm FB}^{\ell}{}^{\rm SM} &=& 0.99571,\, 0.99668,\, 0.99780,
\\
R_\ell^{0\rm}{}^{\rm GHU}/ R_\ell^{0}{}^{\rm SM} &=& 0.99984,\, 0.99989,\,  0.99992,
\\
A_{\rm FB}^b{}^{\rm GHU}/ A_{\rm FB}^b{}^{\rm SM} &=& 0.99769,\, 0.99832,\,  0.99887,
\\
R_b^{\rm GHU} / R_b^{\rm SM} &=& 1.00019,\, 1.00016,\, 1.00014,
\end{eqnarray}
for $\theta_H = 0.115$, $0.0917$ and $0.0737$, respectively.
For $\sqrt{s} = M_Z$, cross section is dominated by the $Z$ boson resonance and 
effects of $Z'$ are very small.
$Z$-boson couplings are very close to the SM value so that the deviation of the cross sections 
from the SM is very tiny. No significant deviations are seen.
In LEP1, the measured $A_{\rm FB}^b$ value deviates from the fit value at nearly $3\sigma$ level.
In GHU, $A_{\rm FB}^b$ is close to the SM value.


In the LEP2 experiment, cross sections of $\bar{q}q$, $\mu^+\mu^-$ and $\tau^+\tau^-$ for twelve different 
collision energies $\sqrt{s}$ ($130\GeV \le \sqrt{s} \le 207\GeV$) were measured.
For the energy 
$120\GeV \lesssim \sqrt{s} \lesssim 207\GeV$,   
$\sigma(\bar{q}q)$ is much larger than the cross sections for lepton final states.
The average of $\sigma^{\rm exp}/\sigma^{\rm SM}(\bar{q}q)$ is $1.0092\pm0.0076$ \cite{Schael:2013ita}. 
In GHU, we obtain $\sigma^{\rm GHU}/\sigma^{\rm SM}(\bar{q}q) = 0.9975$, 
$0.9985$ and $0.9993$ at $\sqrt{s} = 130\GeV$,
and  $0.9882$, $0.9923$ and $0.9953$ at $\sqrt{s} = 207\GeV$ for $\theta_H = 0.115$, $0.0917$ 
and $0.0737$, respectively.
Using the ratios $\sigma^{\rm exp}/\sigma^{\rm SM}$ we perform the $\chi^2$-test for the $\sigma^{\rm model}/\sigma^{\rm SM}$.
When $\sigma^{\rm model} = \sigma^{\rm SM}$, the $\chi^2$-value is $\chi^2/\text{d.o.f.} = 7.3/12$.
In GHU ($\sigma^{\rm model} = \sigma^{\rm GHU}$),
$\chi^2/\text{d.o.f.} = 
14.4/12$ ($p$-value is $28\%$) [ $11.2/7$ ($p$-value is $13\%$)],
$12.0/12$ [$8.9/7$ (26\%)] and
$10.6/12$ [$7.6/7$ (37\%)] for
with 12-bin [7-bin] fit for $\theta_H = 0.115$, $0.0917$ and $0.0737$, respectively.
Here in the 7-bin fit we have chosen seven largest energies ($189\GeV \le \sqrt{s} \le 207\GeV$).
In the fit, correlations among the data are taken into account.
GHU  is within the allowed range of the experimental uncertainty.

Recently at LHC with  $36.4$ fb$^{-1}$ $pp$-collision data
$pp \to \mu^+ \mu^-$ invariant mass distribution $d\sigma/dM_{\mu\mu}$ has been obtained\cite{ATLAS2017}.
The number of observed events and the expected number of the Drell-Yang process in the SM are 
$N^{\rm exp}=4$ and $N^{SM}_{\rm DY} = 5.4\pm0.8$ for $1800\GeV \le M_{\mu\mu} \le 3000\GeV$, respectively.
In GHU with $\theta_H = 0.115$, $0.0917$ and $0.0737$ the expected numbers of  events are
 $N^{\rm GHU}_{\rm DY}/N^{\rm SM}_{\rm DY} = 1.8$, $1.1$ and $0.8$ for $1800\GeV \le M_{\mu\mu} \le 3000\GeV$ 
 \cite{Funatsu:2016uvi}, respectively.\footnote{In the ratio the K-factors in numerators and 
 denominators are cancelled.}
For $\theta_H = 0.115$ the GHU prediction deviates from the SM value (the observed number of events) at 1.8-sigma (2.4-sigma) level.
In \cite{Funatsu:2016uvi}, we have evaluated the expected number of events of Drell-Yang process
with $\sqrt{s_{pp}} = 14\TeV$. For $300\text{ fb}^{-1}$ LHC data, we predict for $\theta_H=0.0737$
the excess of Drell-Yang events as $N_{\rm DY}^{\rm GHU}/N_{\rm DY}^{\rm SM}$
= $42/47$, $6.9/3.6$, $2.6/0.4$ and $1.1/0.04$ for bins [2000, 3000], [3000, 4000], [4000,5000] and [5000, 6000], respectively.

At LEP2, experimental values of $\sigma(\mu^+\mu^-)$, $\sigma(\tau^+\tau^-)$ and $A_{\rm FB}^\ell$ 
have rather large statistical errors and no significant deviations of the ratios 
$\sigma^{\rm GHU} / \sigma^{\rm SM}$ 
and $A_{\rm FB}^{\rm GHU}/A_{\rm FB}^{\rm SM}$ from the experimental data for these modes are seen.


The LHC results put the limit $\theta_H \lesssim 0.1$ on GHU.
To explore GHU one has to go to future $e^+e^-$ colliders at higher energies, 
with $250 \text{ GeV} \le \sqrt{s} \lesssim \text{a few } \text{TeV}$ 
\cite{Fan:2014vta, Baer:2013cma,dEnterria:2016sca,CEPC-SPPCStudyGroup:2015csa,Linssen:2012hp}.
Although with such energy $Z'$s cannot be directory produced, 
the effects of interference among $\gamma$, $Z$ and $Z'$s can be seen.
Furthermore, polarized electron and/or positron beams can be produced at  future $e^+e^-$ colliders.
Since right-handed fermions have larger couplings to $Z'$s in GHU, 
right-handed polarized electron beam will be sensitive to the $Z'$s effects.

Following Ref.~\cite{MoortgatPick:2005cw} we define the longitudinal polarization
$P_{e^\pm}$ ($-1 \le P_{e^\pm}\le 1$) so that 
the electron [positron] is purely right-handed when $P_{e^-}=1$ [$P_{e^+}=1$].
When the vector bosons dominate in the mediators, the cross section at the center-of-mass frame is given by
\begin{align}
\frac{d\sigma}{d\cos\theta} &=
\frac{1}{4} \bigg[ (1-P_{e^-})(1+P_{e^+}) \frac{d\sigma_{LR}}{d\cos\theta} 
  + (1+P_{e^-})(1-P_{e^+}) \frac{d\sigma_{RL}}{d\cos\theta} \bigg],
\label{eq:sigma}
\end{align}
where $\sigma_{LR}$ ($\sigma_{RL}$)
is $e_L^-e_R^+ (e_R^- e_L^+) \to f\bar{f}$ scattering cross section.
Hereafter we consider $\sigma(\bar{q}q)$, $A_{\rm FB}(\mu^+\mu^-)$
and $R_\mu \equiv \sigma(\bar{q}q)/\sigma(\mu^+\mu^-)$.
Although these quantities depend on both $P_{e^-}$ and $P_{e^+}$, the dependence is parameterized 
by one effective polarization $P_{\rm eff} = (P_{e^-} - P_{e^+})/ (1 - P_{e^-} P_{e^+})$.
When $\sigma$ is given by \eqref{eq:sigma}, 
$\sigma(P_{\rm eff},0) = \sigma(P_{e^-},P_{e^+})/(1 - P_{e^-}P_{e^+})$ is satisfied so that 
one finds that $O(P_{\rm eff},0) = O(P_{e^-},P_{e^+})$  where
$O = \sigma^{\rm GHU}(\bar{q}q)/\sigma^{\rm SM}(\bar{q}q)$, $A_{FB}$, $R_\mu$.
As typical values one finds $P_{\rm eff} = \pm0.887$ for $(P_{e^-},P_{e^+}) =  (\pm0.8,\mp0.3)$.
In the following study we parameterize the polarization in terms of $P_{\rm eff}$ instead of 
$( P_{e^-}, P_{e^+})$.

\begin{figure}[btp]
\centerline{\includegraphics[width=0.5\linewidth]{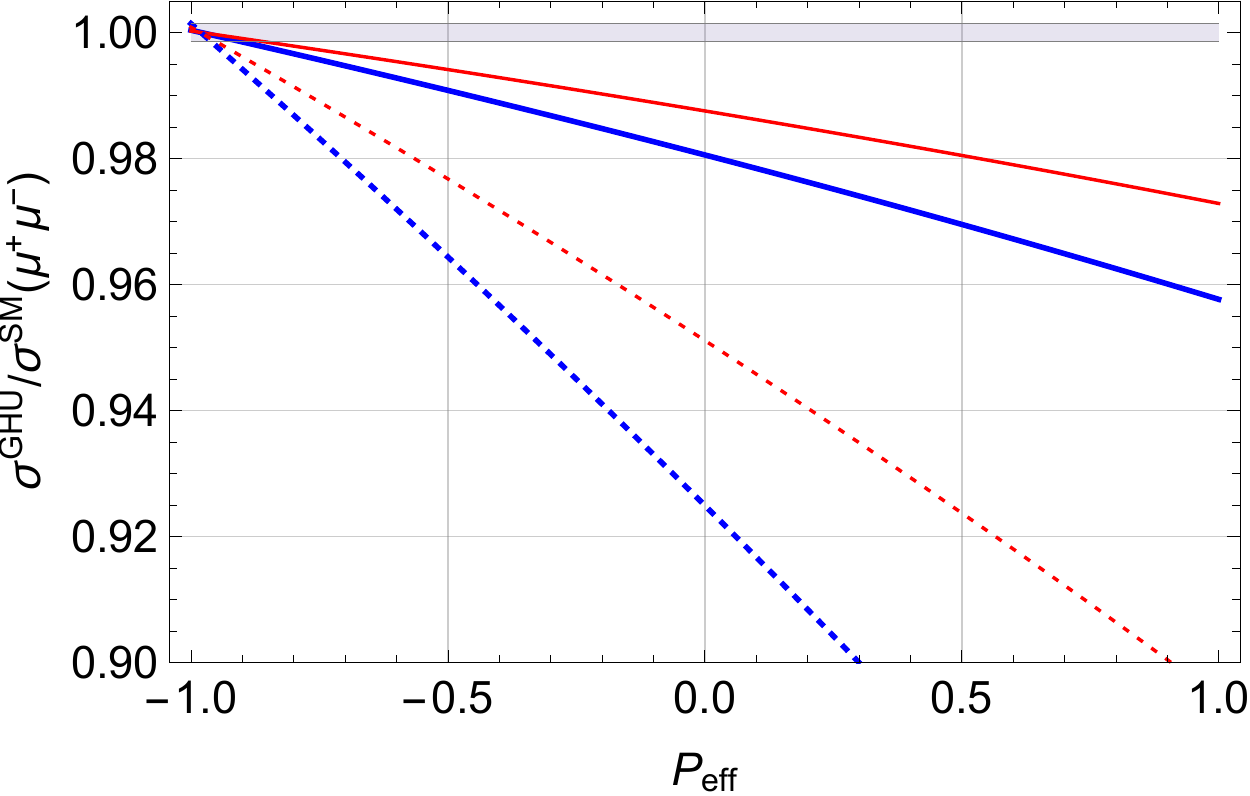}}
\caption{ 
$\sigma(\mu^+\mu^-)$ for the polarized electron and positron beams.
Blue and red lines indicates $\theta_H = 0.0917$ and $0.0737$, respectively.
Solid and dotted lines are for $\sqrt{s} = 250$ GeV and $500$ GeV, respectively.
The gray band indicates  statistical uncertainty at $\sqrt{s}=250\GeV$  
with $250\text{ fb}^{-1}$ data set.}
\label{fig:sigma-ILC}
\end{figure}

At $\sqrt{s} = 250\GeV$ with unpolarized beam (with polarized beam with $P_{\rm eff} = 0.877$),
$\sigma^{\rm SM}(\mu^+\mu^-) = 1.87\text{ pb}$ ($2.16 \text{ pb}$).
In Figure~\ref{fig:sigma-ILC}, the relative cross section 
$\sigma(\mu^+\mu^-)^{\rm GHU}/\sigma(\mu^+\mu^-)^{\rm SM}$ is  
 plotted as a function of $P_{\rm eff}$ at $\sqrt{s} = 250 \GeV$ and $500 \GeV$.
At $\sqrt{s}=250$ GeV,  $\sigma(\mu^+\mu^-)$ in GHU is smaller than the SM value
by  $4.0\%$ [$2.5\%$] for $\theta_H = 0.0917$ [$0.0737$] when $P_{\rm eff} =   0.877$.
At $\sqrt{s}=500$ GeV  with polarization $P_{\rm eff} = 0.877$,  $15\%$ [$9\%$] decrease of $\sigma^{\rm GHU}(\bar{q}q)/\sigma(\bar{q}q)^{\rm SM}$ due to the interference will be observed.
At $\sqrt{s} = 250\GeV$ with $250\text{ fb}^{-1}$ unpolarized $e^+e^-$ beam,
we expect $4.66\times10^5$ $\mu^+\mu^-$ events  in the SM. 
In GHU the expected number of events and statistical significance are estimated to be
 $4.57\times10^5$ [$4.60\times10^5$] and $13.3$ [$8.5$]  for $\theta_H = 0.0917$ [$0.0737$].

Systematic errors in the normalization of the cross sections can be reduced by measuring
\begin{align}
R_{f, \text{RL}} (\overline{P})
=\frac{\sigma( \bar{f}f \, ; \,  P_{e^-} = + \overline{P},  P_{e^+}=0 )}
{\sigma( \bar{f}f \, ; \, P_{e^-} = - \overline{P},  P_{e^+}=0 )} 
\label{defRfRL}
\end{align} 
where the electron beams are polarized with $P_{e^-} = + \overline{P}$  and $- \overline{P}$.
We note that the left-right asymmetry 
$A_{\rm LR}^f \equiv [\sigma_{LR} - \sigma_{RL}] / [\sigma_{LR} + \sigma_{RL}]$ 
is related to $R_{f,RL}$ by
$A_{LR}^f = (\overline{P})^{-1} [1- R_{f,RL}] / [1+ R_{f,RL}]$.
In Table~\ref{tbl:sigmaRL}, the effects of GHU on the $R_{f,\rm RL}$ are tabulated.
GHU predicts a significant deficit in $R_{f, \text{RL}} (\overline{P})$ 
 in the early stage of the ILC experiment.

\begin{table}[htbp]
\caption{$R_{f, \text{RL}} (\overline{P})$  in the SM, and
deviations of $R_{f, \text{RL}} (\overline{P})^{\rm GHU} / R_{f, \text{RL}} (\overline{P})^{\rm SM}$
from unity are tabulated for $\overline{P} = 0.8$.
Statistical uncertainties of $R_{f,\rm RL}^{\rm SM}$ is estimated with $L_{int}$ data for both 
$\sigma( \bar{f}f  ;   P_{e^-} = + \overline{P})$
and $\sigma( \bar{f}f  ;   P_{e^-} = - \overline{P})$, namely with $2 L_{int}$ data in all.}
\label{tbl:sigmaRL}
\vskip 8pt
\centering
%
\begin{tabular}{cc|ccc}
\hline
 $f$ & $\sqrt{s}$~,~ $L_{int}$ & SM & GHU\\
&& $R_{f,RL}^{SM}$ (uncertainty) & $\theta_H=0.0917$ & $\theta_H = 0.0737$ \\
\hline
$\mu$ & $250\GeV$ $250\text{ fb}^{-1}$ & $0.890$ ($0.3\%$) & $-3.4\%$ & $-2.2\%$ \\
      & $500\GeV$ $500\text{ fb}^{-1}$ & $0.900$ ($0.4\%$) & $-13.2\%$ & $-8.6\%$ \\
$b$   & $250\GeV$ $250\text{ fb}^{-1}$ & $0.349$ ($0.3\%$) & $-3.1\%$ &  $-2.1\%$ \\
      & $500\GeV$ $500\text{ fb}^{-1}$ & $0.340$ ($0.5\%$) & $-12.3\%$ & $-8.3\%$ \\      
$t$   & $500\GeV$ $500\text{ fb}^{-1}$ & $0.544$ ($0.4\%$) & $-13.0\%$ & $-8.2\%$ \\
\hline
\end{tabular}
\end{table}

\begin{figure}[htbp]
\centerline{\includegraphics[width=0.5\linewidth]{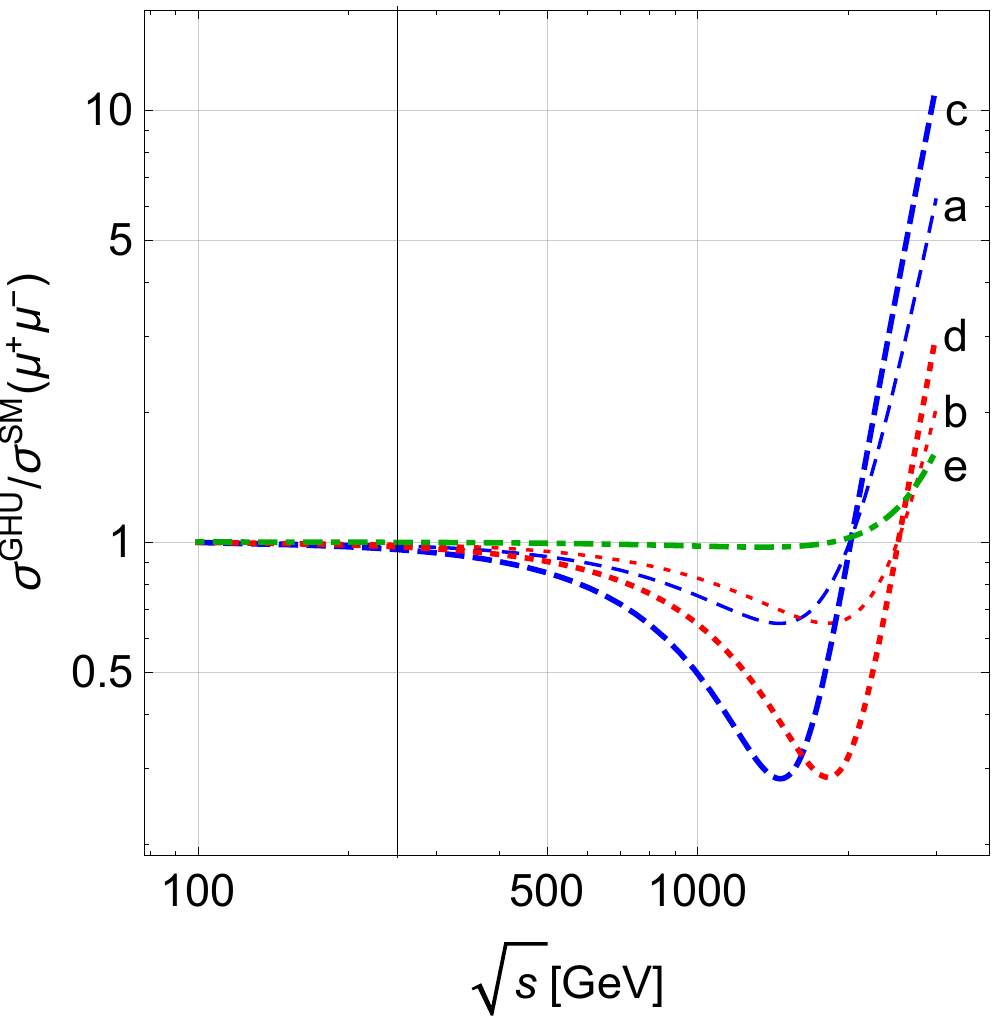}}
\caption{ $\sigma^{\rm GHU}(\mu^+\mu^-)/\sigma^{\rm SM}(\mu^+\mu^-)$ for the polarized electron and positron beams.
``a'', ``c'' and  ``e'' (``b'' and ``d'') are for $\theta_H = 0.0917$ ($0.0737$).
``a'' and ``b'' are for unpolarized beams whereas ``c'' and ``d'' are for polarized beams with 
$P_{\rm eff} = +0.877$.
``e'' is for $P_{\rm eff} = -0.877$.}
\label{fig:sigma-had-HE}
\end{figure}

In Figure~\ref{fig:sigma-had-HE},
$\sigma(\mu^+\mu^-)^{\rm GHU} / \sigma(\mu^+\mu^-)^{\rm SM}$ up to 
$\sqrt{s} = 3 \text{ TeV}$ is displayed. For $1\text{ TeV} \lesssim \sqrt{s} \lesssim 3\text{ TeV}$ 
large deficit is expected for right-handed electron (and/or left-handed positron) beams.
We have also plotted the case (``e'' in the figure) with $P_{\rm eff}= - 0.877$, namely the case with 
left-handed electron (and/or right-handed positron) beams.  
In this case, the interference effect of $Z'$s is hardly seen for $\sqrt{s} < 2 \TeV$.
In all cases the ratios  grow for $\sqrt{s} \gtrsim 3\TeV$ up to the large $Z'$ resonances.

\begin{figure}[htb]
\centerline{\includegraphics[width=0.5\linewidth]{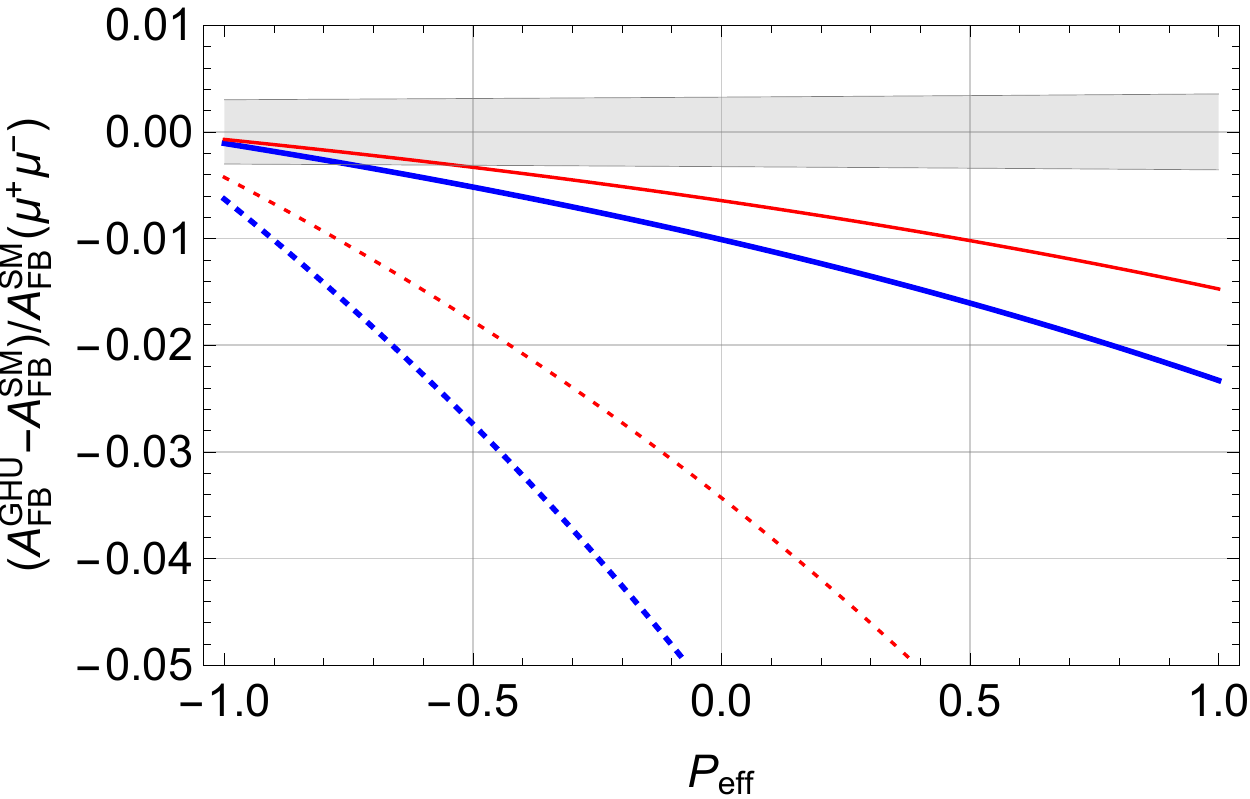}}
\caption{$(A_{\rm FB}^{\rm GHU} - A_{\rm FB}^{\rm SM})/A_{\rm FB}^{\rm SM}(\mu^+\mu^-)$ 
as functions of the effective polarization $P_{\rm eff}$. 
Solid and dotted lines are for $\sqrt{s} = 250\GeV$ and $500$ GeV, respectively.
Blue-thick and red-thin lines correspond to $\theta_H = 0.0917$ and $0.0737$, respectively.
The gray band indicates the  statistical uncertainty at $\sqrt{s}=250\GeV$ with $250\text{ fb}^{-1}$ data.
}\label{fig:FBA-ILC}
\end{figure}

Forward-backward asymmetry (FBA) in the SM at $\sqrt{s} = 250\GeV$ is
$A_{\rm FB}^{\rm SM}(\mu^+\mu^-) = 0.522$ ($0.506$) for $P_{\rm eff}=0$ ($+0.877$).
In Figure~\ref{fig:FBA-ILC}, deviations of $A_{\rm FB}(\mu^+\mu^-)$ from the SM values are plotted 
as functions of $P_{\rm eff}$.   At $\sqrt{s} = 250\GeV$ and for $P_{\rm eff} = 0.877$,
$A_{\rm FB}$ deviates by  $-2.1\%$ ($-1.3\%$) from the SM for $\theta_H = 0.0917$ ($0.0737$).
At $\sqrt{s} = 500\GeV$ and $P_{\rm eff} = 0.877$, $A_{FB}$ deviates by $-12.0\%$ ($-7.4\%$) 
for $\theta_H = 0.0917$ ($0.0737$).
Signals of GHU will be seen at $2\,\sigma$ [$4 \sigma$] level at $\sqrt{s} = 250 \GeV$
 with $250\text{ fb}^{-1}$ unpolarized [polarized] beams.

\begin{figure}[hbt]
\centerline{\includegraphics[width=0.6\linewidth]{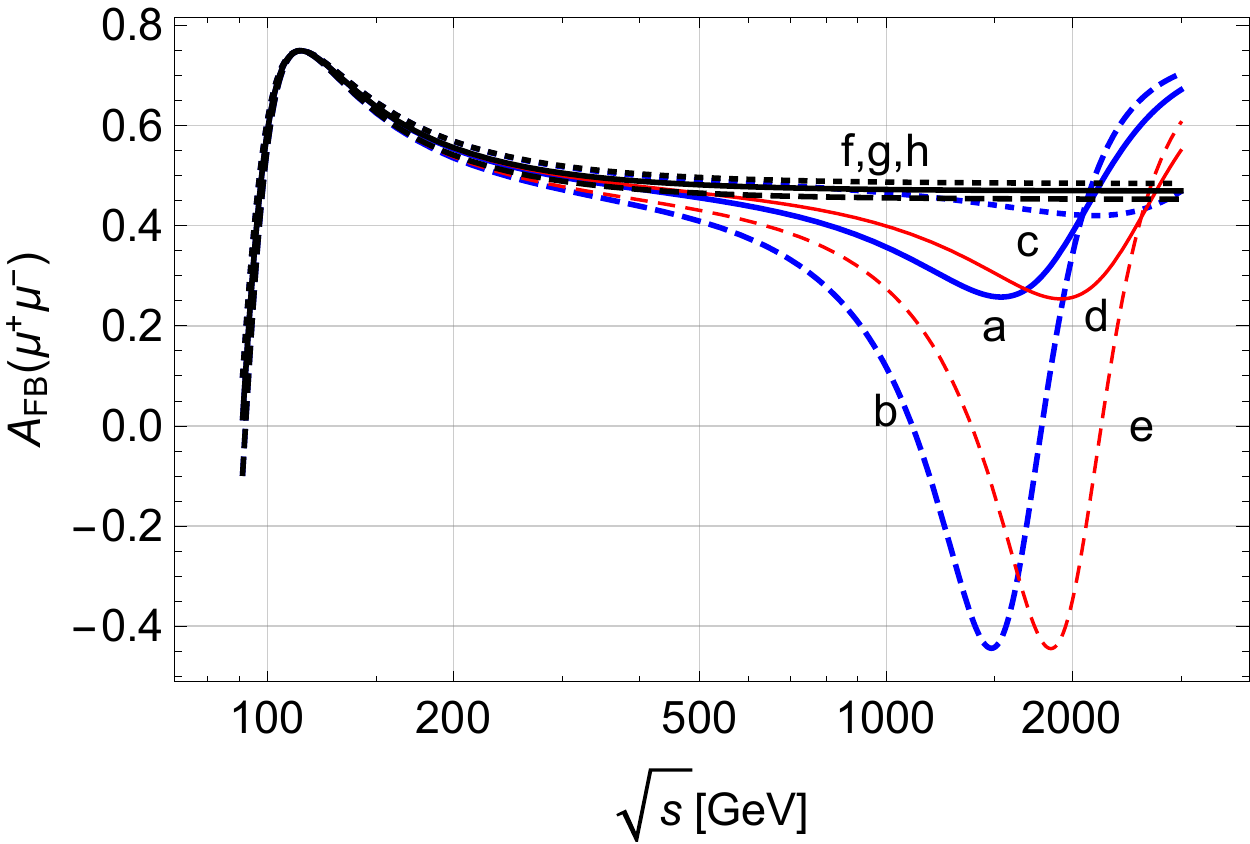}}
\caption{$A_{\rm FB}(\mu^+\mu^-)$ for unpolarized and polarized beams.
``a'', ``b'' and ``c'' (``d'' and ``e'') are for GHU with $\theta_H = 0.0917$ ($0.0737$).
``a'' and ``d'' are for unpolarized beams, 
whereas ``b'' and ``e'' (``c'') are for polarized beams with $P_{\rm eff} = +0.877$ ($-0.877$).
``f''[solid-black], ``g''[dashed-black] and ``h''[dotted-black] correspond to SM with 
unpolarized, $P_{\rm eff} = +0.877$ and $P_{\rm eff}=-0.877$ polarized beams, respectively.
}\label{fig:HE-FBA}
\end{figure}

In Figure~\ref{fig:HE-FBA}, $A_{\rm FB}(\mu^+\mu^-)$ is displayed  up to $\sqrt{s} = 3 \TeV$.
At $\sqrt{s} = 1\sim2 \TeV$, the effect of the interference among $\gamma$, $Z$ and $Z'$ 
becomes maximum. 
In particular for right-handed polarized electron beams very large deviation from the SM is expected.

One can also measure $A_{\rm FB}(\bar{b}b)$, $A_{\rm FB}(\bar{t}t)$.  
They are tabulated in Table.~\ref{tbl:fba}.
We note that $A_{\rm FB}(\bar{b}b)$ and $A_{\rm FB}(\bar{t}t)$  become larger than 
those in the SM, in quite contrast with the $A_{\rm FB}(\mu^+\mu^-)$ case.

\begin{table}[htbp]
\caption{
$(A_{\rm FB}^{q, \rm GHU} - A_{\rm FB}^{q, \rm SM})/A_{\rm FB}^{q, \rm SM}$ ($q = b,t$)}
\label{tbl:fba}
\vskip 7pt
\centering
\begin{tabular}{c|c|c|ccc}
$q\bar{q}$& $\theta_H$ & $\sqrt{s}$ & \multicolumn{3}{c}{$(P_{e^-},\, P_{e^+})$} \\
             &            &            & $(0,\,0)$              & $(+0.8,\, -0.3)$ & $(-0.8,\,+0.3)$ \\
\hline
$b\bar{b}$   & $0.0917$ & $250\GeV$ &  $+0.8\%$ & $+3.3\%$ &$+0.1\%$ \\
             &          & $500\GeV$ &  $+2.9\%$ & $+12.2\%$ & $+0.2\%$ \\
             & $0.0737$ & $250\GeV$ &  $+0.7\%$ & $+3.2\%$ & $+0.1\%$ \\
             &          & $500\GeV$ &  $+2.5\%$ & $+11.2\%$ & $+0.2\%$ \\
\hline
$t\bar{t}$ & $0.0917$ & $500\GeV$ &  $+0.9\%$ & $+4.5\%$ & $+0.1\%$ \\
           & $0.0737$ & $500\GeV$ &  $+1.2\%$  & $+4.2\%$ & $+0.2\%$\\
\end{tabular}
\end{table}

\begin{figure}[htb]
\centerline{\includegraphics[width=0.5\linewidth]{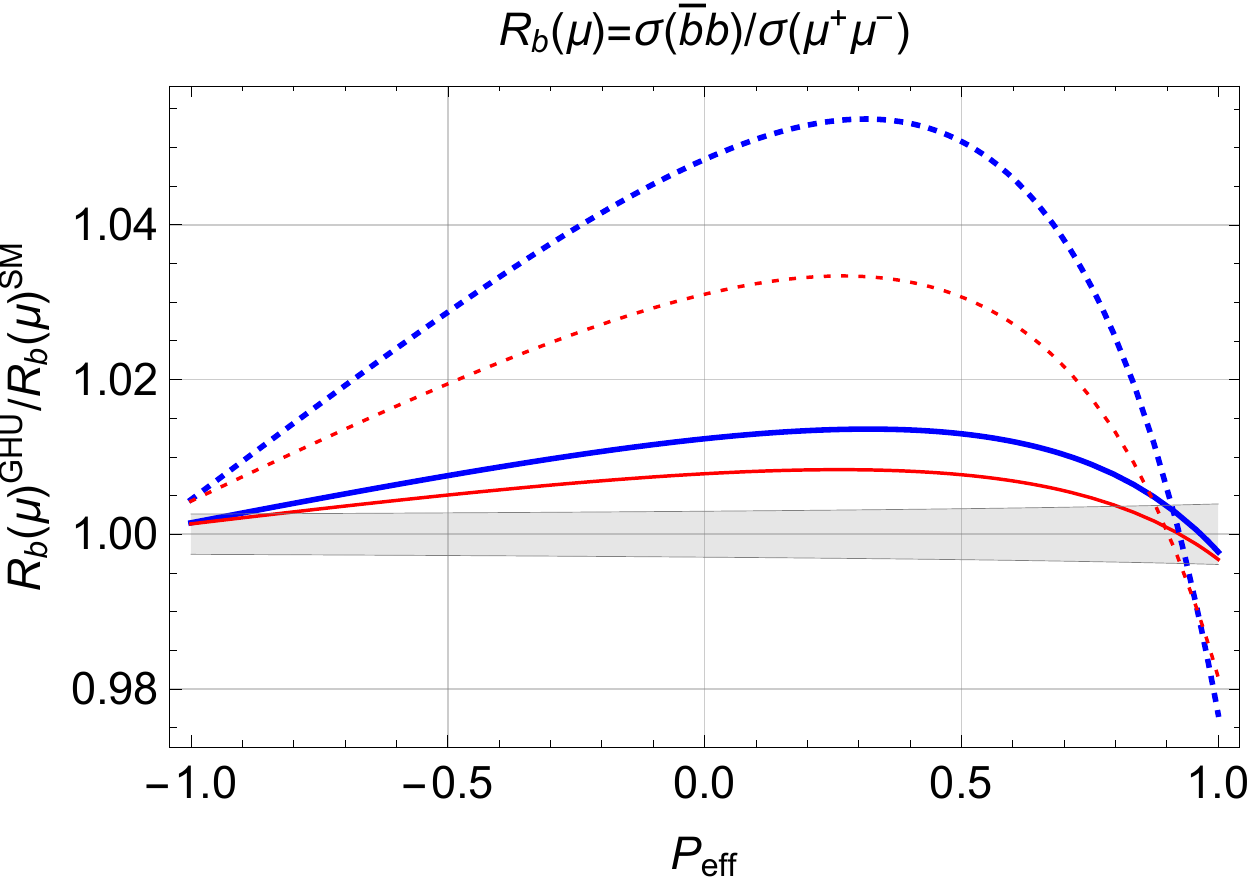}}
\caption{The ratio $R_b(\mu)^{\rm GHU}/R_b(\mu)^{\rm SM}$,  where
$R_b(\mu) \equiv \sigma(\bar{b}b)/\sigma(\mu^+\mu^-)$, is plotted as a function of $P_{\rm eff}$.
Solid and dotted lines are for $\sqrt{s} = 250$ GeV and 500 GeV, respectively.
Blue-thick and red-thin lines are for $\theta_H = 0.0917$ and $0.0737$, respectively.
The gray band indicates the  statistical uncertainty at $\sqrt{s}=250\GeV$ with $250\text{ fb}^{-1}$ data.}\label{fig:Rb}
\end{figure}

The effect of the differences in the couplings of $Z'$ to leptons and quarks can be seen 
in the ratio of the cross sections
$R_b(\mu) \equiv \sigma(\bar{b}b)/\sigma(\mu^+\mu^-)$.\footnote{The $e^+e^- \to \bar{b}b$ 
scattering process contains not only the process mediated by neutral vector bosons,
but also the $W$-fusion process $e^+e^- \to H\nu\bar{\nu}$ followed by $H \to \bar{b}b$.
We have assumed that these processes are efficiently separated and we consider only the vector-boson mediated process.} 
In the SM with unpolarized $e^+e^-$ beams, 
$R_b(\mu)^{\rm SM} = 0.95$, $0.84$ and $0.82$ for $\sqrt{s} = 250\GeV$,
$500\GeV$ and $\infty$, respectively. 
In Figure~\ref{fig:Rb}, deviations of $R_b(\mu)$  from the SM value 
$R_b(\mu)^{\rm GHU} / R_b(\mu)^{\rm SM}$ are plotted as functions of $P_{\rm eff}$.
The excess in $R_b(\mu)$  becomes maximum for $P_{\rm eff} \sim 0.3$.
At $\sqrt{s} = 250 \text{ GeV}$,
$1.1\%$ [$0.8\%$] excess for unpolarized beams and $1.3\%$ [$0.9\%$]
 excess for $P_{\rm eff} = 0.4$ polarized beams for $\theta_H = 0.0917$ [$0.0737$] are expected.
In GHU with $\theta_H \simeq 0.09$, $3\sigma$ deviation is expected with $250\text{ fb}^{-1}$ data. 
At $\sqrt{s} = 500 \text{ GeV}$,  
$5.3\%$ [3.3\%] excess is expected for $P_{\rm eff}=0.4$ polarized beams.
In Table.~\ref{tbl:Rt}, deviation of  $R_{t}(\mu)^{\rm GHU}$ from $R_t(\mu)^{\rm SM}$
is tabulated.  The deviation becomes largest  around $P_{\rm eff} \simeq +0.3$.

\begin{table}[htbp]
\caption{Deviations of the ratio $R_{t}(\mu)^{\rm GHU}/R_t(\mu)^{\rm SM}$ from the unity.}
\label{tbl:Rt}
\vskip 7pt
\centering
\begin{tabular}{c|c|ccc}
$\theta_H$ & $\sqrt{s}$ & \multicolumn{3}{c}{$(P_{e^-},\, P_{e^+})$} \\
           &            & $(0,\,0)$              & $(+0.3,\, 0.0)$ & $(-0.3,0.0)$ \\
\hline
$0.0917$ & $500\GeV$ &  $+2.7\%$ & $+2.7\%$ & $+2.2\%$ \\
$0.0737$ & $500\GeV$ &  $+1.8\%$  & $+1.9\%$ & $+1.5\%$\\
\end{tabular}
\end{table}


In this letter we have studied the effects of the $Z'$ bosons in GHU  in the $e^+e^-$ collider experiments.
At the $Z$ pole ($\sqrt{s} = M_{Z}$), the effects of the $Z'$ bosons  are  small and both cross sections and 
lepton forward-backward asymmetries are consistent with the experiments.
At the energies $130\GeV \le \sqrt{s} \le 207 \GeV$, 
$e^+e^- \to \bar{f}f$ cross sections and forward-backward asymmetry  in GHU 
are found to be consistent with the LEP2 results.
Recent LHC results put the limit $\theta_H \lesssim 0.1$ in GHU.
Large deviations from the SM in $\sigma(\mu^+\mu^-)$, $A_{\rm FB}$,  $R_b(\mu)$ and $R_{f, RL}$
are predicted at higher energies.
In the future $e^+ e^-$ collider experiments, measurements of $\sigma(\mu^+\mu^-)$, $\sigma(\bar{q}q)$, $A_{\rm FB}(\mu^+\mu^-)$ and $R_{f, RL}$
with polarized beams will well discriminate GHU from the SM.
In particular, $\sigma(\mu^+\mu^-)$ measurement, even with unpolarized beams,  
can discriminate the GHU with $\theta_H \simeq 0.09 \, (0.07)$ at 11 (8) times
 of the statistical uncertainty  level  at $\sqrt{s} = 250 \GeV$ with $250\text{ fb}^{-1}$ data.
In the left-right asymmetry $R_{f, RL}$, for which systematic uncertainty is  reduced, 
signals of GHU can be observed at 8 (5) times of the statistical uncertainty  level .
The characteristic dependence of $A_{\rm FB}^\mu$ and $R_b(\mu)$ on the electron-positron 
polarization  can also be used to study the couplings of the $Z'$ bosons to quarks and leptons as well.

The gauge-Higgs unification  is promising. It predicts many signals  in 
$e^+ e^-$ collider experiments.  The left-right asymmetry 
$R_{f, RL} = \sigma(\bar f f ; P_{e^-}= \overline{P})/\sigma(\bar f f ; P_{e^-}= -\overline{P}) $
will exhibit a distinct deviation from the SM in the early stage of 250 GeV ILC with polarized
$e^-$ beams. At 1 TeV ILC or CLIC, clear signals of GHU will be seen in
the forward-backward asymmetry $A_{\rm FB} (\mu^+ \mu^-)$.

\subsection*{Acknowledgments}
The authors would like to thank M.~Tanaka and S.~Kanemura for valuable comments.
This work was supported in part by
the Japan Society for the Promotion of Science, Grants-in-Aid for Scientific Research No 15K05052 (HH and YH).


\renewenvironment{thebibliography}[1]
         {\begin{list}{[$\,$\arabic{enumi}$\,$]}  
         {\usecounter{enumi}\setlength{\parsep}{0pt}
          \setlength{\itemsep}{0pt}  \renewcommand{\baselinestretch}{1.2}
          \settowidth
         {\labelwidth}{#1 ~ ~}\sloppy}}{\end{list}}

\end{document}